\documentclass[12pt]{article}
\usepackage{subfloat}
\pdfoutput = 1
\usepackage{graphics}
\usepackage{graphicx} %Include figure filesusepackage{graphicx} %Include figure files
\begin{document}
\date{Today}
%%%%%%%%%%%%%%%%%%%%
\title{{\bf{\Large  Higher dimensional holographic superconductors in Born-Infeld electrodynamics with backreaction }}}
%%%%%%%%%%%%%%%%%%%%

\author{
{\bf {\normalsize Debabrata Ghorai}$^{a}$
\thanks{debanuphy123@gmail.com, debabrataghorai@bose.res.in}},\,
{\bf {\normalsize Sunandan Gangopadhyay}$^{b,c}
$\thanks{sunandan.gangopadhyay@gmail.com, sunandan@iucaa.ernet.in}}\\
$^{a}$ {\normalsize  S.N. Bose National Centre for Basic Sciences,}\\{\normalsize JD Block, 
Sector III, Salt Lake, Kolkata 700098, India}\\[0.2cm]
$^{b}$ {\normalsize Department of Physics, West Bengal State University, Barasat, India}\\
$^{c}${\normalsize Visiting Associate in Inter University Centre for Astronomy \& Astrophysics,}\\
{\normalsize Pune, India}\\[0.1cm]
}
\date{}

\maketitle

\begin{abstract}
{\noindent In this paper, we analytically investigate the properties of holographic superconductors in higher dimensions
in the framework of Born-Infeld electrodynamics taking into account the backreaction of the spacetime 
using the Sturm-Liouville eigenvalue method.
In the background of pure Einstein and Gauss-Bonnet gravity, based on a perturbative approach, we obtain the relation between the critical temperature and the charge density. 
Higher value of the backreaction and Born-Infeld parameters result
in a harder condensation to form in both cases. 
The analytical results are found to agree with the existing numerical results. We also derive an expression for the condensation operator in $d$-dimensions which yields the critical exponent to be $1/2$. }
\end{abstract}
\vskip 1cm

%%%%%%%%%%%%%%%%%%%%%%%%%%%%%%%%% Introduction %%%%%%%%%%%%%%%%%%%%%%%%%%%%%%%%%%%%%%%%%%%%%%%%%%%%
\section{Introduction}

It is well known that weakly coupled superconductors can be described with great accuracy by the
BCS theory of superconductivity \cite{bcs}, which is based on the fact that the interaction between electrons resulting from the virtual exchange of phonons is attractive when the energy difference between the states of the electrons is less than the energy of the phonon. However, progress in this field in the last few decades has made it clear that this microscopic theory fails in understanding the pairing mechanism in materials (like high $T_c$ cuprates) which are strongly coupled. The pairing mechanism and the normal state of the system before the onset of superconductivity has been eluding theorists for a long time and hence forces one to think about other alternative theories. One such alternative theory is provided by the AdS/CFT correspondence.

The AdS/CFT correspondence proposed by Maldacena \cite{adscft1} has drawn the attention of condensed matter theorists because of its remarkable ability to address issues in strongly interacting systems by exploiting results obtained in a weakly coupled gravitational system. The duality \cite{adscft1}-\cite{adscft4} provides an exact correspondence between gravity theory in a $(d+1)$-dimensional AdS spacetime and a conformal field theory (CFT) sitting on the $d$-dimensional boundary of this spacetime. In recent years, it has provided a holographic description of properties of $s$-wave superconductors, namely, the holographic superconductor phase transition \cite{hs1}-\cite{hs3}, the holographic Fermi liquid \cite{nw1},\cite{nw2} and the holographic insulator/superconductor phase transition \cite{nw3}. This model consisting a black hole and a complex scalar field minimally coupled to an abelian gauge field is observed to form a scalar hair below a certain critical temperature $T_c$ due to the breaking of a local $U(1)$ gauge symmetry near the black hole event horizon \cite{hs1}-\cite{hs5}.

A number of studies have been carried out on various holographic superconductor models based on the framework of Maxwell electrodynamics \cite{hs6}-\cite{hs18}
as well as non-linear electrodynamics \cite{hs19}-\cite{sgm}, namely, Born-Infeld electrodynamics \cite{hs19}. For example, in \cite{hs6}, a holographic dual description of a superconductor had been provided via second order phase transition in which the condensate determined the energy gap formed due to frequency dependent conductivity below a critical temperature. In \cite{siop}, analytical techniques had been employed to investigate the properties of holographic superconductors (in particular in elucidating the nature of the ground state) in the framework of Maxwell electrodynamics. Further, the Mermin-Wagner theorem suggests that the phase transition may be affected by higher curvature corrections. Investigations in this direction led to the introduction of a new analytic method in \cite{hs8}, the so called matching method, which is based on the solution to the field equation near the horizon and near the asymptotic region and then matching the two solutions at some intermediate point. Several properties like critical exponents \cite{hs9a}, various condensates \cite{hs16} had been studied in the framework of Gauss-Bonnet gravity \cite{hs9b} (which takes into account the effect of higher curvature corrections) analytically as well as numerically \cite{hs9c}-\cite{hs18}.
The analytical study of properties of holographic
superconductors in Einstein gravity in the framework of Born-Infeld electrodynamics was first carried out in \cite{sgdc1}
using the Sturm-Liouville (SL) eigenvalue approach. The physical motivation of looking at the leading order corrections coming from the
Born-Infeld coupling parameter is to investigate the effects due to higher derivative corrections of gauge fields on the order parameter condensation.

The Gauss-Bonnet (GB) gravity \cite{deser}-\cite{cai} has attracted a lot of attention among gravity theories with higher curvature corrections.
As mentioned earlier, the study of curvature corrections in $(3+1)$-dimensional holographic superconductors has been of some interest because of the Mermin-Wagner theorem which forbids continuous  symmetry breaking in $(2+1)$-dimensions. In \cite{hs8}, \cite{hs24} analytical studies in
GB gravity have been carried out using the matching as well as the SL method thereby revealing that higher curvature corrections
make the formation of the scalar hair harder. However, these studies are based on the probe limit which neglects the back reactions
of matter fields on the spacetime metric \cite{nell}-\cite{sgplb}.

In this paper, we shall study the properties of holographic superconductors in higher dimensions in the framework of 
Born-Infeld electrodynamics away from the probe limit using the SL eigenvalue approach. In particular we obtain the relation between the critical temperature and the charge density of holographic superconductors in $d$-dimensions in the framework of Einstein and GB gravity and then study the $d=5$ case. We have presented the plots of $T_{c} $ vs. $\rho^{1/3} $ which clearly show that the inclusion of any of the parameters, namely, the Born-Infeld parameter, the GB coupling parameter and the back reaction of the matter fields on the spacetime metric makes the scalar hair formation harder. Our analytical results agree with studies carried out in the literature using the matching method \cite{yao}. We further investigate the relation between the condensation operator and the critical temperature. We compute an expreesion for the condensation operator in $d$-dimensions and then study the effects of the Born-Infeld and Gauss-Bonnet parameters in the presence of back reactions in $d=5$ dimensions.

%%%%%%%%%%%%%%%%%%%%%%%%%%%%%%%%%%%%%%%%%%%%%%%%%%%%%%%%%%%%%%%%%%%%%%%%%%%%%%%%%%%%%%%%%%

%%%%%%%%%%%%%%%%%%%%%%%%%%%%%%%%%%%%%%%%%%%%%%%%%%%%%%%%%%%%%%%%%%%%%%%%%%%%%%%%%%%%%%%%%% 

This paper is organized as follows. In section 2, we provide the basic holographic set up for the holographic superconductors in the background of a $d$-dimensional electrically charged black hole in anti-de Sitter spacetime. In section 3, taking into account the effect of the Born-Infeld electrodynamics and the backreaction of the matter fields on the spacetime metric in Einstein and Gauss-Bonnet gravity, we compute the critical temperature in terms of a solution to the Sturm-Liouville eigenvalue problem. In section 4, we analytically obtain an expression for the condensation operator in $d$-dimension near the critical temperature. We conclude finally in section 5.

%%%%%%%%%%%%%%%%%%%%%%%%%%%%%%%%%%%%%%%%%%%%%%%%%%%%%%%%%%%%%%%%%%%%%%%%%%%%%%%%%%%%%%%%%%%

%%%%%%%%%%%%%%%%%%%%%%%%%%%%%  Section 2     %%%%%%%%%%%%%%%%%%%%%%%%%%%%%%%%%%%%%%%%%%%%%%

\section{Basic formalism }
Our basic starting point is to write down the action for the formation of scalar
hair on an electrically charged black hole in $d$-dimensional anti-de Sitter spacetime. This reads
\begin{eqnarray}
S=\int d^{d}x \frac{\sqrt{-g}}{2\kappa^{2}} \left( R -2\Lambda +\frac{\alpha}{2}(R^{2} - 4 R^{\mu \nu} R_{\mu \nu} + R^{\mu \nu \lambda \rho } R_{\mu \nu \lambda \rho }) + 2 \kappa^2 \mathcal{L}_{matter} \right)
\label{ac1}
\end{eqnarray}
where $\Lambda=-(d-1)(d-2)/(2L^2)$ is the cosmological constant, $\kappa^2 = 8\pi G_{d} $ is the $d $-dimensional Newton's gravitational constant and $\alpha$ is the Gauss-Bonnet coupling parameter.\\
The matter Lagrangian density is denoted by $\mathcal{L}_{matter}$ and reads
\begin{eqnarray}
\mathcal{L}_{matter}= \frac{1}{b}\left(1-\sqrt{1+ \frac{b}{2}F^{\mu \nu} F_{\mu \nu}}\right) - (D_{\mu}\psi)^{*} D^{\mu}\psi-m^2 \psi^{*}\psi
\label{ac2}
\end{eqnarray}
where $F_{\mu \nu}=\partial_{\mu}A_{\nu}-\partial_{\nu}A_{\mu}$; ($\mu,\nu=0,1,2,3,4$) is the field strength tensor, 
$D_{\mu}\psi=\partial_{\mu}\psi-iqA_{\mu}\psi$ is the covariant derivative,  $A_{\mu}$ and $ \psi $ represent the gauge field and scalar field respectively.\\
We now assume that the plane-symmetric black hole metric with back reaction can be written in the form
\begin{eqnarray}
ds^2=-f(r)e^{-\chi(r)}dt^2+\frac{1}{f(r)}dr^2+ r^2 h_{ij} dx^{i} dx^{j}
\label{m1}
\end{eqnarray}
where $ h_{ij} dx^{i} dx^{j}$ denotes the line element of a $(d-2) $-dimensional hypersurface with zero curvature. The Hawking temperature of this black hole, which is interpreted as the temperature of the conformal field theory on the boundary, is given by \cite{chen}
\begin{eqnarray}
T_{H} = \frac{f^{\prime}(r_{+}) e^{-\chi(r_{+})/2}}{4\pi}
\label{gzx1}
\end{eqnarray} 
where $ r_{+} $ is the radius of the horizon of the black hole. 

\noindent The ansatz for the gauge field and the scalar field is now chosen to be \cite{hs6}
\begin{eqnarray}
A_{\mu} = (\phi(r),0,0,0)~,~\psi=\psi(r).
\label{vector}
\end{eqnarray}
The above ansatz implies that the black hole possesses only electric charge.

\noindent The equations of motion for the metric and matter fields calculated with this ansatz read
\begin{eqnarray}
&&\left(1-\frac{2 \alpha f(r)}{r^2}\right) f^{\prime}(r) + \frac{(d-3)f(r)}{r} - \frac{(d-1)r}{L^2} + \frac{2 \kappa^2 r}{d-2} \nonumber\\
&\times & \left[f(r)\psi^{\prime}(r)^2 + \frac{q^2 \phi^2(r) \psi^2(r) e^{\chi(r)}}{f(r)} + m^2 \psi^2(r) \frac{q^2 \phi^2(r) \psi^2(r) e^{\chi(r)}}{f(r)} + \frac{1}{b}\left((1- b\phi^{\prime}(r)^{2})^{-\frac{1}{2}} -1\right)\right]=0\nonumber \\
\label{e2}
\end{eqnarray}
\begin{eqnarray}
\left(1-\frac{2 \alpha f(r)}{r^2}\right)\chi^{\prime}(r) + \frac{4 \kappa^2 r}{d-2}\left(\psi^{\prime}(r)^2 + \frac{q^2 \phi^2(r) \psi^2(r) e^{\chi(r)}}{f(r)^2}\right) = 0
\label{e02}
\end{eqnarray}
\begin{eqnarray}
\phi^{\prime \prime}(r) + \left(\frac{d-2}{r}+ \frac{\chi^{\prime} (r)}{2}\right) \phi^{\prime}(r) - \frac{d-2}{r} b e^{\chi(r)}\phi^{\prime}(r)^{3} - \frac{2 q^2 \phi(r) \psi^{2}(r)}{f(r)}(1 - b e^{\chi(r)} \phi^{\prime}(r)^{2})^\frac{3}{2} = 0
\label{e1}
\end{eqnarray}
\begin{eqnarray}
\psi^{\prime \prime}(r) + \left(\frac{d-2}{r}- \frac{\chi^{\prime} (r)}{2} + \frac{f^{\prime}(r)}{f(r)}\right)\psi^{\prime}(r) + \left(\frac{q^2 \phi^{2}(r) e^{\chi(r)}}{f(r)^2}- \frac{m^{2}}{f(r)}\right)\psi(r) = 0
\label{e01}
\end{eqnarray}
where prime denotes derivative with respect to $r$. 
The fact that $\kappa\neq0$ takes into account the backreaction of the spacetime. Without any loss of generality, this limit also enables one to choose $q=1$  since the rescalings $\psi\rightarrow \psi/q$, $\phi\rightarrow \phi/q$ and $\kappa^2 \rightarrow q^2 \kappa^2 $ can be performed \cite{betti}.

\noindent We now proceed to solve the non-linear equations (\ref{e2})-(\ref{e01}). In order to do this we need
to fix the boundary conditions for $\phi(r)$ and $\psi(r)$ at the black
hole horizon $r=r_+$ (where $f(r=r_+)=0$ with $e^{-\chi(r=r_+)}$ finite) and at the spatial infinity
($r\rightarrow\infty$). For the matter fields to be regular, we require $\phi(r_+)=0$ and $\psi(r_{+})$ to be finite at the horizon. 

\noindent Near the boundary of the bulk, we can set $e^{-\chi(r\rightarrow\infty)}\rightarrow1$, so that the spacetime becomes a
Reissner-Nordstr\"{o}m-anti-de Sitter black hole. The matter fields there obey \cite{hs8}
\begin{eqnarray}
\label{bound1}
\phi(r)&=&\mu-\frac{\rho}{r^{d-3}}\\
\psi(r)&=&\frac{\psi_{-}}{r^{\Delta_{-}}}+\frac{\psi_{+}}{r^{\Delta_{+}}}
\label{bound2}
\end{eqnarray}
where
\begin{eqnarray}
\label{bound1a}
\Delta_{\pm}&=&\frac{(d-1)\pm\sqrt{(d-1)^2+4m^2 L^2}}{2}~.
%\label{bound3calculations for other set of values %%%%%%%%%%%%%%%%%%%%%%a}
\end{eqnarray}
The parameters $\mu$ and $\rho$ are dual to the chemical potential and charge density of the conformal field theory on the boundary.
We choose $\psi_{-}=0$, so that $\psi_{+}$ is dual to the expectation value of the condensation operator $J$ at the boundary.

\noindent Under the change of coordinates $z=\frac{r_{+}}{r}$,  the field equations (\ref{e2})-(\ref{e01}) become
\begin{eqnarray}
&&\left(1-\frac{2 \alpha z^2 f(z)}{r^2_{+}}\right)f^{\prime}(z) - \frac{(d-3)f(z)}{z} + \frac{(d-1)r^2_{+}}{L^2 z^3} - \frac{2 \kappa^2 r^2_{+}}{(d-2)z^3} \nonumber\\ &\times& \left[\frac{z^4}{r^2_{+}} f(z)\psi^{\prime}(z)^2 + \frac{\phi^2(z) \psi^2(z) e^{\chi(z)}}{f(z)} + m^2 \psi^2(z) \frac{\phi^2(z) \psi^2(z) e^{\chi(z)}}{f(z)} + \frac{1}{b}\left((1- \frac{b z^4}{r^2_{+}}\phi^{\prime}(z)^{2})^{-\frac{1}{2}} -1\right)\right]=0\nonumber\\
\label{e1a}
\end{eqnarray}
\begin{eqnarray}
\left(1-\frac{2 \alpha z^2 f(z)}{r^2_{+}}\right)\chi^{\prime}(z) - \frac{4 \kappa^2 r^2_{+}}{(d-2)z^3}\left(\frac{z^4}{r^2_{+}}\psi^{\prime}(z)^2 + \frac{\phi^2(z) \psi^2(z) e^{\chi(z)}}{f(z)^2}\right) = 0
\label{e1ab}
\end{eqnarray}
\begin{eqnarray}
\phi^{\prime \prime}(z) + \left(\frac{\chi^{\prime}(z)}{2} -\frac{d-4}{z}\right) \phi^{\prime}(z) + \frac{d-2}{r^2_{+}} b e^{\chi(z)}\phi^{\prime}(z)^{3} z^3 - \frac{2r^2_{+} \phi(z) \psi^{2}(z)}{f(z) z^4}\left(1 -\frac{b z^4 e^{\chi(z)}}{r^2_{+}} \phi^{\prime}(z)^{2}\right)^\frac{3}{2} = 0 \nonumber\\
\label{e1aa}
\end{eqnarray}
\begin{eqnarray}
\psi^{\prime \prime}(z) + \left(\frac{f^{\prime}(z)}{f(z)} - \frac{d-4}{z}- \frac{\chi^{\prime}(z)}{2}\right)\psi^{\prime}(z) + \frac{r^2_{+}}{z^4} \left(\frac{\phi^{2}(z) e^{\chi(z)}}{f(z)^2}- \frac{m^{2}}{f(z)}\right)\psi(z) =0 
\label{e1bb}
\end{eqnarray}
where prime now denotes derivative with respect to $z$. These equations are to be solved in the interval $(0, 1)$, where $z=1$ is the horizon and $z=0$ is the boundary.
The boundary condition $\phi(r_+)=0$ now translates to $\phi(z=1)=0$.

%%%%%%%%%%%%%%%%%%%%%%%%%%%%%%%%%%%%%%%%%%%%%%%%%%%%%%%%%%%%%%%%%%%%%%%%%%%%

%%%%%%%%%%%%%%%%%%%%%%%%  Section 3 %%%%%%%%%%%%%%%%%%%%%%%%%%%%%%%%%%%%%%%%

\section{Relation between critical temperature($T_{c}$) and charge density($\rho$)}
With the basic formalism in place, in this section we shall proceed to investigate the relation between the critical temperature and the charge density. To begin with, we first 
need to obtain a solution of eq.(\ref{e1aa}).

\noindent At the critical temperature $T_c$, $\psi=0$, hence eq.(\ref{e1ab}) reduces to 
\begin{eqnarray}
\label{bk1}
\chi^{\prime}(z)= 0.
\label{bk2}
\end{eqnarray}
Near the boundary of the bulk, we can set $e^{-\chi(r\rightarrow\infty)} \rightarrow 1 $, i.e. $\chi(r\rightarrow\infty)=0$
which in turn implies $\chi(z)=0$ from eq.(\ref{bk2}). The field equation (\ref{e1aa}) therefore reduces to
\begin{eqnarray}
\label{metric1}
\phi^{\prime \prime}(z) -\frac{d-4}{z}\phi^{\prime}(z) + \frac{(d-2)bz^3}{r^2_{+(c)}} \phi^{\prime}(z)^3=0.
\end{eqnarray}
To solve this non-linear differential equation, we take recourse to a perturbative technique developed in \cite{sgdc1}. 
When $b=0$, the above equation becomes  
\begin{eqnarray}
\phi^{\prime \prime}(z) -\frac{d-4}{z}\phi^{\prime}(z) =0.
\label{sol}
\end{eqnarray}
Using the asymptotic behaviour of $\phi(z)$ (eq.(\ref{bound1})), the solution of eq.(\ref{sol}) reads
\begin{eqnarray}
\phi(z)|_{b=0}= \lambda r_{+(c)} (1-z^{d-3})
\label{de1}
\end{eqnarray}
where
\begin{eqnarray}
\lambda=\frac{\rho}{r_{+(c)}^{d-2}}~.
\label{lam}
\end{eqnarray}
To solve eq.(\ref{metric1}), we put the solution for $\phi(z)$ with $b=0$ (i.e. $\phi(z)|_{b=0}$) in the non-linear term of 
eq.(\ref{metric1}). This leads to
\begin{eqnarray}
\phi^{\prime \prime}(z) -\frac{d-4}{z} \phi^{\prime}(z) - b\lambda^3 r_{+(c)} (d-2)(d-3)^3 z^{3(d-3)} =0.
\label{de2}
\end{eqnarray}    
Using the asymptotic boundary condition(\ref{bound1}), the solution of the above equation 
upto first order in the Born-Infeld parameter $b$ reads 
\begin{eqnarray}
\phi(z) = \lambda r_{+(c)}\left\{ (1-z^{d-3}) - \frac{b(\lambda^2|_{b=0}) (d-3)^3}{2(3d-7)} (1-z^{3d-7})\right\}
\label{de3} 
\end{eqnarray}
where we have used the fact that $b\lambda^2= b(\lambda^2|_{b=0}) + \mathcal{O}(b^2)$ \cite{sgdc1}, $\lambda^2|_{b=0}$ being the value of $\lambda^2$ for $b=0$. It is reassuring to note that the above result agrees with solution of $\phi (z)$ obtained in \cite{sgdc1} for $d=4$.

%%%%%%%%%%%%%%%%%%%%%%%%%%%%%%%%%%%%%%%%%%%%%%%%%%%%%%%%%%%%%%%%%

%%%%%%%%%%%%%%%%%%%%%%%% Sub-section 3.1 %%%%%%%%%%%%%%%%%%%%%%%%
  
\subsection{Backreaction effect in Einstein gravity}
For Einstein gravity $\alpha=0$, the eq.(\ref{e1a}) at $T=T_{c}$ becomes
\begin{eqnarray}
f'(z) -\frac{d-3}{z}f(z) +\frac{(d-1)r^2_{+(c)}}{L^2 z^3} - \frac{2 \kappa^2 r^2_{+(c)}}{(d-2)z^3}\frac{1}{b}\left([1- \frac{b z^4}{r^2_{+(c)}}\phi^{\prime}(z)^{2}]^{-\frac{1}{2}} -1\right)=0.
\label{de4}
\end{eqnarray}
Dropping terms of the order of $b \kappa^2$, eq.(\ref{de4}) reduces to 
\begin{eqnarray}
f'(z) -\frac{d-3}{z}f(z) +\frac{(d-1)r^2_{+(c)}}{L^2 z^3} -\frac{\kappa^2 z}{d-2}\phi^{\prime}(z)^2 =0 
\label{de5}
\end{eqnarray}
Substituting $\phi(z)|_{b=0}$ (eq.(\ref{de1})) in the above equation leads to 
\begin{eqnarray}
f'(z) -\frac{d-3}{z}f(z) +\frac{(d-1)r^2_{+(c)}}{L^2 z^3} -\frac{\kappa^2 \lambda^2 r^2_{+(c)} (d-3)^2}{d-2} z^{2d-7} =0.
\label{de6} 
\end{eqnarray}
The solution of the metric from eq.(\ref{de6}) subject to the condition $f(z=1)=0$ reads
\begin{eqnarray}
f(z)&=&r_{+(c)}^{2}\left\{\frac{1}{L^2 z^2}-\left(\frac{1}{L^2}+\frac{d-3}{d-2}\kappa^2 \lambda^2\right)z^{d-3}+\frac{d-3}{d-2}\kappa^2 \lambda^2 z^{2(d-3)}\right\}.
\label{metr2}
\end{eqnarray} 
In the rest of the analysis we shall set $L=1$. Eq.(\ref{metr2}) therefore reads 
\begin{eqnarray}
f(z)= \frac{r^2_{+(c)}}{z^2} g_{0}(z)
\label{de7}
\end{eqnarray}
where 
\begin{eqnarray}
g_{0}(z)= 1- \left(1+\frac{d-3}{d-2}\kappa^2 \lambda^2\right)z^{d-1}+ \frac{d-3}{d-2}\kappa^2 \lambda^2 z^{2(d-2)}.
\label{metr33}
\end{eqnarray} 

%%%%%%%%%%%%%%%%%%%%%%%%%%%%%%%%%%%%%%%%%%%%%%%%%%%%%%%%%%%%%%%%%%%%%%%%%

\noindent Now we find that as $T\rightarrow T_c$, eq.(\ref{e1bb}) for the field $\psi$ approaches the limit
\begin{eqnarray}
\psi''(z)+ \left(\frac{g'_{0}(z)}{g_{0}(z)}-\frac{d-2}{z}\right)\psi'(z)+ \left( \frac{\phi^{2} (z)}{g^{2}_{0} (z) r^{2}_{+(c)}} - \frac{m^2}{g_{0}(z) z^2}\right)\psi(z)=0
\label{e001}
\end{eqnarray}
where $\phi(z)$ now corresponds to the solution in eq.(\ref{de3}). In the above equation, we
shall also consider the fact that $\kappa^2_{i} \lambda^2 =\kappa^2_{i}(\lambda^2|_{\kappa_{i-1}}) +\mathcal{O}(\kappa^4)$ which 
in turn implies that $g_{0}(z)$ reads like
\begin{eqnarray}
g_0(z) =1- \left(1+\frac{d-3}{d-2}\kappa^2_{i} (\lambda^2|_{\kappa_{i-1}})\right)z^{d-1}+ \frac{d-3}{d-2}\kappa^2_{i} 
(\lambda^2|_{\kappa_{i-1}}) z^{2(d-2)}.
\end{eqnarray}
Near the boundary, we define \cite{siop}
\begin{eqnarray}
\psi(z)=\frac{<J>}{r^{\Delta_{+}}_{+(c)}} z^{\Delta_{+}} F(z)
\label{sol1}
\end{eqnarray}
where $F(0)=1$ and $J$ is the condensation operator.
Substituting this form of $\psi(z)$ in eq.(\ref{e001}), we obtain
\begin{eqnarray}
F''(z) &+& \left\{\frac{2\Delta_{+} -d+2}{z} +\frac{g'_{0}(z)}{g_{0}(z)} \right\}F'(z) +\left\{ \frac{\Delta_{+}(\Delta_{+} -1)}{z^2} +\left(\frac{g'_{0}(z)}{g_{0}(z)}-\frac{d-2}{z}\right)\frac{\Delta_{+}}{z}-\frac{m^2}{g_{0}(z) z^2} \right\}F(z) \nonumber \\
&+& \frac{\lambda^2}{g^{2}_{0}}\left\{ (1-z^{d-3})^2 -\frac{b(\lambda^2|_{b=0})(d-3)^3}{3d-7}(1-z^{d-3})(1-z^{3d-7})\right\}F(z)=0
\label{eq5b}
\end{eqnarray}
to be solved subject to the boundary condition $F' (0)=0$. 

\noindent It is now simple to see that the above equation can be written in the Sturm-Liouville form 
\begin{eqnarray}
\frac{d}{dz}\left\{p(z)F'(z)\right\}+q(z)F(z)+\lambda^2 r(z)F(z)=0
\label{sturm}
\end{eqnarray}
with 
\begin{eqnarray}
p(z)&=&z^{2\Delta_{+} -d+2}g_{0}(z)\nonumber\\
q(z)&=&z^{2\Delta_{+} -d+2}g_{0}(z)\left\{ \frac{\Delta_{+}(\Delta_{+} -1)}{z^2} +\left(\frac{g'_{0}(z)}{g_{0}(z)}-\frac{d-2}{z}\right)\frac{\Delta_{+}}{z}-\frac{m^2}{g_{0}(z) z^2} \right\} \nonumber\\
r(z)&=&\frac{z^{2\Delta_{+} -d+2}}{g_{0}(z)} \left\{ (1-z^{d-3})^2 -\frac{b(\lambda^2|_{b=0}) (d-3)^3}{3d-7}(1-z^{d-3})(1-z^{3d-7})\right\}~. 
\label{i1}
\end{eqnarray}
The above identification enables us to write down an equation for the eigenvalue $\lambda^2$ which minimizes the expression 
\begin{eqnarray}
\lambda^2 &=& \frac{\int_0^1 dz\ \{p(z)[F'(z)]^2 - q(z)[F(z)]^2 \} }
{\int_0^1 dz \ r(z)[F(z)]^2}~.
\label{eq5abc}
\end{eqnarray}
We shall now use the trial function for the estimation of $\lambda^{2}$
\begin{eqnarray}
F= F_{\tilde\alpha} (z) \equiv 1 - \tilde\alpha z^2
\label{eq50}
\end{eqnarray}
Note that $F$ satisfies the conditions $F(0)=1$ and $F'(0)=0$.

\noindent Using eq.(\ref{gzx1}) and eq.(s)(\ref{de7}, \ref{metr33}), 
we get the relation between the critical temperature and the charge density
\begin{eqnarray}
T_{c}=\frac{1}{4\pi}\left[ (d-1)- \frac{(d-3)^2}{(d-2)}\kappa^2_{i} (\lambda^2|_{\kappa_{i-1}})\right]\left(\frac{\rho}{\lambda}\right)^{\frac{1}{d-2}}.
\label{de9} 
\end{eqnarray}
The above result holds for a $d$-dimensional holographic superconductor and is one of the main results in this paper. It is to be noted that the effect of the BI coupling parameter $b$ in the critical temperature $T_{c}$ comes through the eigenvalue $\lambda$. 
In the rest of our analysis, we shall set $d=5$ and $m^2=-3$. The choice for $m^2$ yields $\Delta_{+}=3$ from eq.(\ref{bound1a}). 
Eq.(s)(\ref{de9}, \ref{i1}) therefore becomes
\begin{eqnarray}
\label{de12}
\label{nwe}
T_{c}&=&\frac{1}{\pi}\left[ 1-\frac{1}{3}\kappa^2_{i} (\lambda^2|_{\kappa_{i-1}}) \right]\left(\frac{\rho}{\lambda}\right)^{\frac{1}{3}}\\
p(z)&=&z^3 \left\{ 1-z^4\left(1+\frac{2}{3}\kappa^2_{i} (\lambda^2|_{\kappa_{i-1}})\right)+\frac{2}{3}\kappa^2_{i} 
(\lambda^2|_{\kappa_{i-1}}) z^6 \right\} \nonumber \\
q(z)&=& -9 z^5 \left( 1+\frac{2}{3}\kappa^2_{i} (\lambda^2|_{\kappa_{i-1}})\right) + 10\kappa^2_{i} (\lambda^2|_{\kappa_{i-1}}) z^7 \nonumber \\
r(z)&=&\frac{z^3\left\{ (1-z^{2})^2 -b(\lambda^2|_{b=0}) (1-z^{2})(1-z^{8})\right\}}{1-z^4\left(1+\frac{2}{3}\kappa^2_{i} 
(\lambda^2|_{\kappa_{i-1}})\right)+\frac{2}{3}\kappa^2_{i} (\lambda^2|_{\kappa_{i-1}}) z^6}~.
\label{de13}
\end{eqnarray}
With the backreaction parameter $\kappa=0$ and Born-Infeld parameter $b=0$, the trial function (\ref{eq50}) 
and eq.(\ref{de13}) leads to
\begin{eqnarray}
\lambda_{\tilde\alpha}^2 = \frac{2(18-27\tilde\alpha+14{\tilde\alpha}^2 )}{6(3-4\ln2)+16(2-3\ln2)\tilde\alpha + (17-24\ln2)
{\tilde\alpha}^2} 
\label{est2}
\end{eqnarray}
which attains its minimum at $\tilde\alpha \approx 0.7218$. 
The critical temperature can now be computed from eq.(\ref{de12}) and reads 
\begin{eqnarray}
T_c =\frac{1}{\pi(\lambda|_{\tilde\alpha=0.7218})^{1/3}}\rho^{1/3}\approx 0.1962\rho^{1/3} 
\label{eqTc}
\end{eqnarray}
which is in very good agreement with the numerical 
$T_c = 0.1980\rho^{1/3}$ \cite{hs16}.\\

\noindent Now in order to include the effect of the Born-Infeld parameter $b$, we set $b=0.01$ 
and rerun the above analysis to get the  value of $\lambda^2$ for $b=0.01$ 
\begin{eqnarray}
\lambda_{\tilde\alpha}^2 = \frac{1.500-2.250\tilde\alpha +1.6667{\tilde\alpha}^2}
{0.0371037-0.0316927\tilde\alpha + 0.00845841{\tilde\alpha}^2} 
\label{est1}
\end{eqnarray}
which attains its minimum at $\tilde\alpha \approx 0.7540$. 
The critical temperature therefore reads 
\begin{eqnarray}
T_c =\frac{1}{\pi\lambda_{\tilde\alpha=0.7540}^{1/3}}\rho^{1/3}\approx 0.1850\rho^{1/3} 
\label{eqTc1}
\end{eqnarray}
which is in very good agreement with the exact $T_c = 0.1910\rho^{1/3}$ \cite{hs24}. 

\noindent Setting $b=0.02$ yields
\begin{eqnarray}
\lambda_{\tilde\alpha}^2 = \frac{1.500-2.250\tilde\alpha +1.6667{\tilde\alpha}^2}
{0.0173545-0.0104244\tilde\alpha + 0.00173068{\tilde\alpha}^2}
\label{z1}
\end{eqnarray} 
which attains its minimum at $\tilde\alpha \approx 0.8201$. Hence the critical temperature reads
\begin{eqnarray}
T_c =\frac{1}{\pi\lambda_{\tilde\alpha=0.8201}^{1/3}}\rho^{1/3}\approx 0.1694\rho^{1/3}
\label{z2}
\end{eqnarray}
which is in good agreement with the exact $T_c = 0.1851\rho^{1/3}$ \cite{hs24}. A comparison of the analytical and numerical results for the critical temperature and the charge density in Einstein gravity with backreaction parameter $\kappa=0$ is presented in Table 1.\\
\begin{table}
\caption{For backreaction parameter $\kappa=0$}
\centering
\begin{tabular}{c c c c c}
\hline
$ b$ & $\tilde\alpha $ & $\lambda^{2}_{SL}$&$(T_{c}/\rho^{1/3})|_{SL}$ & $(T_{c}/\rho^{1/3})|_{numerical}$ \\
\hline
0.0 & 0.7218 & 18.23 & 0.1962 & 0.1980 \\
0.01 & 0.7540 & 25.91 & 0.1850 & 0.1910 \\
0.02 & 0.8201 & 44.08 & 0.1694 & 0.1851\\ 
\hline
\end{tabular}
\end{table}

\noindent Now we shall proceed to include the effect of backreaction ($\kappa\neq 0 $) in the above analysis. To do this, we shall increase
the value of $\kappa$ in steps of $0.05$. To begin with, we set $b=0$ and the backreaction parameter $\kappa=0.05$. 
Rerunning the above procedure using eq.(s)(\ref{eq50}, \ref{de13})  leads to 
\begin{eqnarray}
\lambda_{\tilde\alpha}^2 = \frac{1.48861-2.22721\tilde\alpha +1.15401{\tilde\alpha}^2}
{0.057139-0.0533188\tilde\alpha + 0.0153081{\tilde\alpha}^2}~.
\label{z3}
\end{eqnarray}
This attains its minimum at $\tilde\alpha \approx 0.7195$. The critical temperature therefore reads
\begin{eqnarray}
T_c =\frac{1}{\pi}\frac{(1-\frac{1}{3}\kappa^2\lambda_{\kappa=0.0}^2)}{\lambda_{\tilde\alpha=0.7195}^{1/3}}\rho^{1/3}\approx 0.1934\rho^{1/3}
\label{z4}
\end{eqnarray}
which is in good agreement with the exact $T_c = 0.1953\rho^{1/3}$ \cite{chen}. We repeat our calculations of $T_{c}$ for the 
same value of $\kappa$ but with different values of $b$. The critical temperature reads
$T_{c}=0.1825\rho^{1/3}$ and $T_{c}=0.1672\rho^{1/3} $ for $b=0.01,~0.02$ respectively. Next we repeat the same analysis for $\kappa= 0.10$ and $\kappa= 0.15$. 

\noindent Figure 1 shows the plot of $T_{c}$ vs. $\rho$ for Einstein holographic superconductors for different choice of parameters $\kappa,~b$. The plots clearly show that the condensation becomes harder to form as the values of the backreaction parameter $\kappa$ and BI coupling parameter $b$ are increased. \\

%%%%%%%%%%%%%%%%%%%%%%%%%%%%%%%%%%%%%%%%%%%%%%%%%%%%%%%%%%%%%%%%%%%%%%%%%%%%%%%%%%%%%%%%%%%%%%%%%%%%%%%%%%%%%%%%%%
%%%%%%%%%%%%%%%%%%%%%%%%%%%%%%%%%%%%%%% %%%%%%%%%%%%%%%

%%%%%%%%%%%%%%%%%%%%%%%%%%%%%%%%%%%%%%%%%%%%%%%%%%%%%%%%%%%%%%%%%%%%%%%%%%%%%%%%%%%%%%%%%%%%%%%%%%%%
%%%%%%%%%%%%%%%%%%%%%%%%%%%%%%%%%%%%%%%%%%%%%%%%%%%%%%%%%%%%%%%%%%%%%%%%%%%%%%%%%%%%%%%%%%%%%%%%%%%%

\noindent In Table 2, we present our analytical values obtained by the SL eigenvalue approach for different sets of values of $b $ and $\kappa$. In figure 1, we show the effect of backreaction as well as BI coupling parameters on the critical temperature $(T_{c})$.
\begin{table}[htb]
\caption{The analytical results for the critical temperature and the charge density with backreaction and Born-Infeld parameter in Einstein gravity}   
\centering                          
\begin{tabular}{c c c c c c c}            
\hline                       
$\kappa$& $b$ & $\tilde{\alpha} $ & $\lambda^{2}_{SL}$  & $(T_{c}/\rho^{1/3})|_{SL}$ \\
\hline
 & 0.0 & 0.7195 & 18.11 & 0.1934  \\ 
0.05 & 0.01 & 0.7525 & 25.68 & 0.1825    \\
 & 0.02 & 0.8203 & 43.46 & 0.1672    \\     
\hline 
 & 0.0 & 0.7122 & 17.75 & 0.1852  \\  
0.10 & 0.01& 0.7455 & 25.02 & 0.1751 \\  
 & 0.02 & 0.8148 & 41.73 & 0.1608  \\ 
\hline 
 & 0.0& 0.6995 & 17.16 & 0.1718  \\   
0.15 & 0.01& 0.7345 & 23.99 &  0.1634  \\   
 & 0.02& 0.8024 & 39.07 & 0.1506  \\
\hline                 
\end{tabular}\label{E1}  
\end{table}

\begin{figure}[t]
%\begin{subfigure}
\includegraphics[scale=0.32]{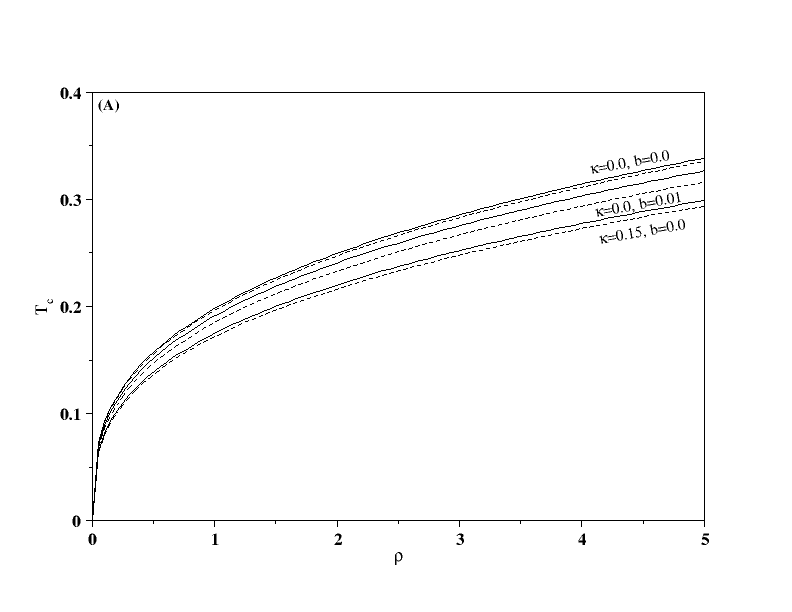}  
%\end{subfigure}
%\begin{subfigure}
\includegraphics[scale=0.4]{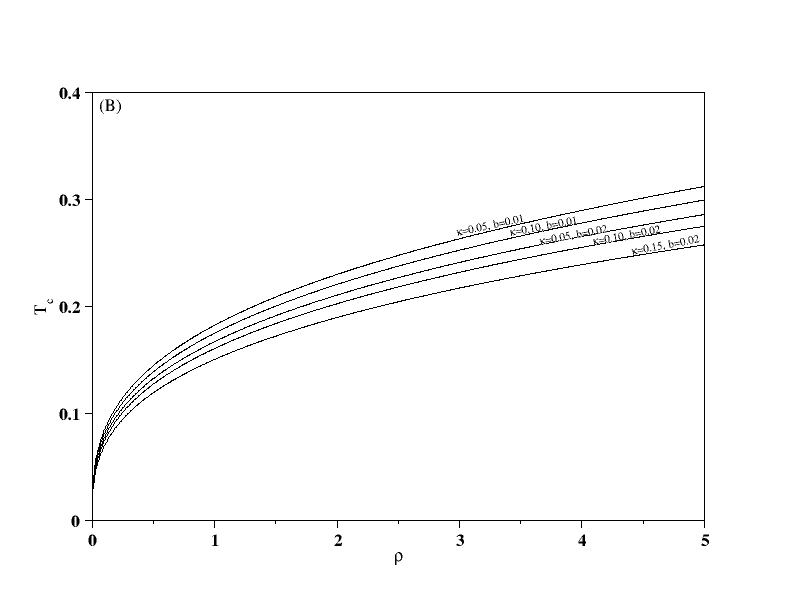}
%\end{subfigure}
%\includegraphics[scale=.4]{pic/debu1}
\caption{$T_{c}$ vs. $\rho$ plot for Einstein holographic superconductors for different choice of parameters $\kappa,~b$.
(A) The continuous curves correspond to numerical values whereas the dotted curves correspond to analytic values for $(\kappa=0,~b=0),~(\kappa=0,~b=0.01),~(\kappa=0.15,~b=0)$.
(B) All curves correspond to analytic values for $(\kappa=0.05,~b=0.01),~(\kappa=0.10,~b=0.01),~(\kappa=0.05,~b=0.02), ~(\kappa=0.10,~b=0.02), ~(\kappa=0.15,~b=0.02)$ (upper to lower).}
\end{figure}

%%%%%%%%%%%%%%%%%%%%%%%%%%%%%%%%%%%%%%%%%%%%%%%%%%%%%%%%%%%%%%%%%%%%%%%%%
%%%%%%%%%%%%%%%%%%%%%%%%%%%%%%%%%%%%%%%%%%%%%%%%%%%%%%%%%%%%%%%%%%%%%%%%%

%%%%%%%%%%%%%%%%%%%%%%%%%%%%%       Sub-section 3.2                %%%%%%%%%%%%%%%%%%%%%%%%%%%%%%%%%%%%%%%%%%%%

\subsection{Backreaction effect in Gauss-Bonnet gravity}
In this subsection, we study the relation between the critical temperature and the charge density taking into account the effect of the Gauss-Bonnet coupling parameter $\alpha$. It is to be noted that since $\kappa$ and $b$ are very small, 
hence we shall neglect $\mathcal{O}(b\kappa^2)$ and higher order terms in our analysis. 

\noindent In this case, using eq.(\ref{bk2}), eq.(\ref{e1a}) (with $\alpha\neq0$) reduces to
\begin{eqnarray}
\left(1-\frac{2\alpha z^2}{r^{2}_{+(c)}} f(z) \right)f'(z) -\frac{d-3}{z}f(z) +\frac{(d-1)r^2_{+(c)}}{L^2 z^3} -\frac{\kappa^2 z}{d-2}\phi^{\prime}(z)^2 =0~.
\label{e10bbb}
\end{eqnarray}
Since we are not concerned with terms of the order of $b\kappa^2$,  
we substitute $\phi(z)|_{b=0}$ (eq.(\ref{de1})) in eq.(\ref{e10bbb}). The metric equation then becomes
\begin{eqnarray}
f'(z) -\frac{d-3}{z}f(z) +\frac{(d-1)r^2_{+(c)}}{L^2 z^3} -\frac{\kappa^2 \lambda^2 r^2_{+(c)} (d-3)^2}{d-2} z^{2d-7} =\frac{2\alpha z^2}{r^{2}_{+(c)}} f(z) f'(z).
\label{de15}
\end{eqnarray}
To solve this non-linear equation, we once again employ the perturbative technique. First we consider $\alpha=0$ for which we know the solution, namely, $f(z)|_{\alpha=0}=\frac{r^2_{+(c)}}{z^2} g_{0}(z)$. To solve for $\alpha \neq 0$, we substitute $f(z)|_{\alpha=0}$ and $f'(z)|_{\alpha=0}$ in the right hand side of eq.(\ref{de15}). 
The solution of the above equation upto first order in the Gauss-Bonnet coupling parameter $\alpha$ therefore reads
\begin{eqnarray}
f(z)&=&\frac{r_{+(c)}^{2}}{z^2}\left\{g_{0}(z)+2\alpha g_{1}(z)\right\}
\label{metr20z}
\end{eqnarray} 
where 
\begin{eqnarray}
g_{1}(z)&=&\frac{2}{d-1} -z^{d-1} -(d-5)z^{d-1}\log z+\frac{d-3}{d-1}z^{2(d-1)} \nonumber \\
&+&\left\{ \frac{2(d-4)}{d-2}z^{2(d-2)} -\frac{3(d-3)^2}{2(d-2)^2}z^{3d-5}+ \frac{2(d-3)^2}{(d-2)(d-1)}z^{2(d-1)}\right\}\kappa^2_{i}(\lambda^2|_{\kappa_{i-1}})\nonumber\\ &+& z^{d-1}\left\{\frac{77-95d+39d^2 -5d^3}{2(d-1)(d-2)^2} -\frac{(d-3)(d-5)}{d-2}\log z\right\} \kappa^2_{i}(\lambda^2|_{\kappa_{i-1}})~.
\label{metr41}
\end{eqnarray} 
Once again substituting the form $\psi(z)$ near the boundary (defined in eq.(\ref{sol1}))
%\begin{eqnarray}
%\psi(z)=\frac{<J>}{r^{\Delta_{+}}_{+(c)}} z^{\Delta_{+}} F(z)~.
%\label{sol1a}
%\end{eqnarray}
in eq.(\ref{e001}), we obtain
\begin{eqnarray}
F''(z) &+& \left\{\frac{2\Delta_{+} -d+2}{z} +\frac{g'_{0}(z)+2\alpha g'_{1}(z)}{g_{0}(z)+ 2\alpha g_{1}(z)} \right\}F'(z) \nonumber \\
&+&\left\{ \frac{\Delta_{+}(\Delta_{+} -1)}{z^2} +\left(\frac{g'_{0}(z)+2 \alpha g'_{1}(z)}{g_{0}(z)+ 2\alpha g_{1}(z)}-\frac{d-2}{z}\right)\frac{\Delta_{+}}{z}-\frac{m^2}{(g_{0}(z)+ 2\alpha g_{1}(z)) z^2} \right\}F(z) \nonumber\\
&+& \frac{\phi^2(z)|_{b=0}}{r^2_{+(c)}\left(g_{0}(z)+ 2\alpha g_{1}(z)\right)
^2} F(z)=0~
\label{eq5ba}
\end{eqnarray}
to be solved subject to the boundary condition $F' (0)=0$. 

\noindent The above equation can once again be put in the Sturm-Liouville form (\ref{sturm})
with 
\begin{eqnarray}
p(z)&=&z^{2\Delta_{+} -d+2}\left(g_{0}(z)+ 2\alpha g_{1}(z) \right)\nonumber\\
q(z)&=&z^{2\Delta_{+} -d+2}\left(g_{0}(z)+ 2\alpha g_{1}(z) \right)\left\{ \frac{\Delta_{+}(\Delta_{+}-d +1)}{z^2} +\left(\frac{g'_{0}(z)+2\alpha g'_{1}(z)}{g_{0}(z)+2\alpha g_{1}(z)}\right)\frac{\Delta_{+}}{z}-\frac{m^2}{\left(g_{0}(z)+ 2\alpha g_{1}(z) \right) z^2} \right\} \nonumber\\
r(z)&=&\frac{z^{2\Delta_{+} -d+2}}{\left(g_{0}(z)+ 2\alpha g_{1}(z) \right)} \left\{ (1-z^{d-3})^2 -\frac{\lambda^2\mid_{b=0} b(d-3)^3}{3d-7}(1-z^{d-3})(1-z^{3d-7})\right\}~. 
\label{i1aa}
\end{eqnarray}
With the above identification, we can once again proceed to find the minimum value of the eigenvalue $\lambda^2$ as
in the earlier section. 

\noindent Once again using eq.(\ref{gzx1}) and eq.(s)(\ref{metr20z}, \ref{metr41}), we get the relation between the critical temperature and the charge density.\\
It is to be noted that the expression for the critical temperature in GB gravity is identical to the corresponding expression in Einstein gravity (\ref{de9}). This is because $g_{1}^{\prime}(z)$ vanishes at $z=1$. However, their numerical values will be different since in GB gravity, the eigenvalues $\lambda$ will be affected by the GB coupling parameter $\alpha$. \\

%%%%%%%%%%%%%%%%%%%%%%%%%%%%%%%%%%%%%%%%%%%%%%%%%%%%%%%%%%%%%%%%%%%%%%%%%%%%%

% At $T\rightarrow T_{c}$, the critical temperature reads 
%\begin{eqnarray}
%T_{c}=-\frac{f'(z=1)}{r_{+(c)}}=-\left[ g'_{0}(z=1)+2\alpha g'_{1}(z=1)\right]r_{+(c)}
%\end{eqnarray}
%Using eq.(\ref{metr33}) and eq.(\ref{metr41}), we get same result as get previously because $g'_{1}(z=1)=0$ i.e. eq.(\ref{de9}). But $T_{c}$ will effected by Gauss-Bonnet parameter $\alpha$ through $\lambda^2$ values.\\

%%%%%%%%%%%%%%%%%%%%%%%%%%%%%%%%%%%%%%%%%%%%%%%%%%%%%%%%%%%%%%%%%%%%%%%%%%%%%

\noindent Setting $d=5$ and $m^2=-3$, eq.(s)(\ref{de9}, \ref{i1aa}) becomes
\begin{eqnarray}
T_{c}&=&\frac{1}{\pi}\left[ 1-\frac{1}{3}\kappa^2_{i} (\lambda^2|_{\kappa_{i-1}}) \right]\left(\frac{\rho}{\lambda}\right)^{\frac{1}{3}}\\
p(z)&=&z^3 \left\{ 1-z^4\left(1+\frac{2}{3}\kappa^2_{i} (\lambda^2|_{\kappa_{i-1}})\right)+\frac{2}{3}\kappa^2_{i} 
(\lambda^2|_{\kappa_{i-1}}) z^6 \right\} \nonumber \\
&+& 2\alpha z^3\left\{\frac{1}{2}(1+z^8) -z^4 -\frac{2}{3} \kappa^2_{i} (\lambda^2|_{\kappa_{i-1}})\left( z^4 -z^6 -z^8 +z^{10}\right)\right\} \nonumber \\
q(z)&=& -9 z^5 \left( 1+\frac{2}{3}\kappa^2_{i} (\lambda^2|_{\kappa_{i-1}})\right) + 10\kappa^2_{i} (\lambda^2|_{\kappa_{i-1}}) z^7 \nonumber \\
&+& \alpha \left\{ 21z^9 -18z^5 -3z -4\kappa^2_{i} (\lambda^2|_{\kappa_{i-1}})\left(3z^5 -5z^7 -7z^9 +9z^{11}\right) \right\}\nonumber \\
r(z)&=&\frac{z^3\left\{ (1-z^{2})^2 -b(\lambda^2|_{b=0}) (1-z^{2})(1-z^{8})\right\}}{1-z^4\left(1+\frac{2}{3}\kappa^2_{i} \lambda^2|_{\kappa_{i-1}}\right)+\frac{2}{3}\kappa^2_{i} \lambda^2|_{\kappa_{i-1}} z^6 +2\alpha\left\{\frac{1}{2}(1+z^8) -z^4 -\frac{2}{3} \kappa^2_{i} \lambda^2|_{\kappa_{i-1}}\left( z^4 -z^6 -z^8 +z^{10}\right)\right\}}.\nonumber\\
\label{a1}
\end{eqnarray}  
 
\noindent To estimate $\lambda^2$, we first set $\alpha=-0.1$, $\kappa=0$, $b=0$ and once again use the trial function (\ref{eq50}) to obtain
\begin{eqnarray}
\lambda_{\tilde\alpha}^2 = \frac{1.26-2.00\tilde{\alpha}+1.07143{\tilde\alpha}^2 }{0.0613835-0.0565523\tilde\alpha + 0.0160771{\tilde\alpha}^2} 
\label{a2}
\end{eqnarray}
which attains its minimum at $\tilde\alpha \approx 0.7305$. 
The critical temperature therefore reads 
\begin{eqnarray}
T_{c} =\frac{1}{\pi\lambda_{\tilde\alpha=0.7305}^{1/3}}\rho^{1/3}\approx 0.208\rho^{1/3} 
\label{eqTc1a}
\end{eqnarray}
which is in very good agreement with the exact $T_c = 0.209\rho^{1/3}$ \cite{liu}. 

\noindent Next we include effect of backreaction.
We calculate $\lambda^2$ for $\kappa=0.01,~ b=0$ which attains its minimum at $\tilde\alpha \approx 0.7345$. The critical temperature therefore reads 
\begin{eqnarray}
T_c &=&\frac{1}{\pi}\frac{(1-\frac{1}{3}\kappa^2\lambda_{\kappa=0.0}^2)}{\lambda_{\tilde\alpha=0.7345}^{1/3}}\rho^{1/3}\approx 0.2077\rho^{1/3}
\label{a3}
\end{eqnarray}
which is in very good agreement with the exact $T_c = 0.2089\rho^{1/3}$ \cite{yao}.

%(ref. jhep05(2013)101). 

%Similarly we have calculated the relation between $T_{c}$ and $\rho$ for $\kappa=0.02$ which is shown in the  table 3.
\begin{table}
\caption{For $\alpha=-0.1$,~ $b=0$}
\centering
\begin{tabular}{c c c c c}
\hline
$\kappa $ & $\tilde\alpha $ & $\lambda^{2}_{SL}$&$(T_{c}/\rho^{1/3})|_{SL}$ & $(T_{c}/\rho^{1/3})|_{numerical}$ \\
\hline
0.0 & 0.7305 & 12.940 & 0.2078 & 0.2090 \\
0.01 & 0.7345 & 12.937 & 0.2077 & 0.2089 \\
0.02 & 0.7302 & 12.930 & 0.2074 & 0.2087 \\ 
\hline
\end{tabular}
\end{table}
%We set $\alpha=0.0001$, $\kappa=0$, $b=0$ and obtain
%\begin{eqnarray}
%\lambda_{\tilde\alpha}^2 = \frac{1.50025-2.25025\tilde{\alpha}+1.16676{\tilde\alpha}^2 }{0.0568487-0.0529577\tilde\alpha + 0.0151848{\tilde\alpha}^2} 
%\end{eqnarray}
%which attains its minimum at $\tilde\alpha \approx 0.7218$. 
%The critical temperature therefore reads 
%\begin{eqnarray}
%T_{c} =\frac{1}{\pi\lambda_{\tilde\alpha=0.7218}^{1/3}}\rho^{1/3}\approx 0.1962\rho^{1/3} 
%\label{eqTc1a}
%\end{eqnarray}
%which is in very good agreement with the exact $T_c = 0.1962\rho^{1/3}$ (ref.jhep05(2012)156). Similarly we have calculated for same %value of $\kappa,~ \alpha$ with different value of $b=0.01,~ 0.02$ in the table 4.
\begin{table}
\caption{For $\alpha=0.0001$,~$\kappa=0$}
\centering
\begin{tabular}{c c c c c}
\hline
$b $ & $\tilde\alpha $ & $\lambda^{2}_{SL}$&$(T_{c}/\rho^{1/3})|_{SL}$ & $(T_{c}/\rho^{1/3})|_{numerical}$ \\
\hline
0.0 & 0.7218 & 18.2358 & 0.1962 & 0.1962 \\
0.01 & 0.7565 & 25.8432 & 0.1851 & 0.1910 \\
0.02 & 0.8211 & 44.105 & 0.1693 & 0.1851 \\ 
\hline
\end{tabular}
\end{table}
%Now in order to include the effect of backreaction, we set $\kappa=0.05,~ 0.10,~ 0.15$. And we set $b=0.0,~ 0.01,~ 0.02$ for each $\kappa$'s value, which is shown in the table 5. 
\begin{table}
\caption{The analytical results for the critical temperature and the charge density with backreaction and Born-Infeld parameter in Gauss-Bonnet gravity ($\alpha=0.0001$)}   
\centering                          
\begin{tabular}{c c c c c c c}            
\hline                       
$\kappa$& $b$ & $\tilde{\alpha} $ & $\lambda^{2}_{SL}$  & $(T_{c}/\rho^{1/3})|_{SL}$ \\
\hline
 & 0.0 & 0.7205 & 18.12 & 0.1934  \\ 
0.05 & 0.01 & 0.7505 & 25.75 & 0.1824    \\
 & 0.02 & 0.8275 & 42.93 & 0.1675    \\     
\hline 
 & 0.0 & 0.7125 & 17.75 & 0.1852  \\  
0.10 & 0.01& 0.7454 & 25.04 & 0.1751 \\  
 & 0.02 & 0.8135 & 41.70 & 0.1608  \\ 
\hline 
 & 0.0& 0.7011 & 17.16 & 0.1718  \\   
0.15 & 0.01& 0.7318 & 24.00 &  0.1633  \\   
 & 0.02& 0.8025 & 39.10 & 0.1505  \\
\hline                 
\end{tabular}\label{E5}  
\end{table}

%Now we take $\alpha=0.1,~ \kappa=0,~ b=0$, so we obtain
%\begin{eqnarray}
%\lambda_{\tilde\alpha}^2 = \frac{1.74-2.50\tilde{\alpha}+1.2619{\tilde\alpha}^2 }{0.052994-0.049849\tilde\alpha + 0.014403{\tilde\alpha}^2} 
%\end{eqnarray}
%which attains its minimum at $\tilde\alpha \approx 0.7080$. 
%The critical temperature therefore reads 
%\begin{eqnarray}
%T_{c} =\frac{1}{\pi\lambda_{\tilde\alpha=0.7080}^{1/3}}\rho^{1/3}\approx 0.187\rho^{1/3} 
%\label{eqTc1a}
%\end{eqnarray}
%which is in very good agreement with the exact $T_c = 0.185\rho^{1/3}$ (ref.jhep0910(2009)010).

\noindent In the tables below, we present the analytical results obtained by the SL approach for different sets of 
values of $\alpha$, $\kappa$ and $b$. 

\noindent In figure 2, the plot of $T_{c}$ vs. $\rho$ is shown for holographic superconductors in the framework of Gauss-Bonnet gravity for different choice of parameters $\kappa,~b$. The plots clearly show that the condensation becomes harder to form as the values of the backreaction parameter $\kappa$, BI coupling parameter $b$ and the GB parameter $\alpha$ are increased.

\begin{table}
\caption{The analytical results for the critical temperature and the charge density with backreaction and Born-Infeld parameter in Gauss-Bonnet gravity ($\alpha=0.1$)}   
\centering                          
\begin{tabular}{c c c c c c c}            
\hline                       
$\kappa$& $b$ & $\tilde{\alpha} $ & $\lambda^{2}_{SL}$  & $(T_{c}/\rho^{1/3})|_{SL}$ \\
\hline
 & 0.0 & 0.7080 & 24.18 & 0.1872 \\
0.0 & 0.01 & 0.7665 & 39.96 & 0.1722 \\
 & 0.02 & 0.9375 & 103.31 & 0.1470\\ 
\hline
 & 0.0 & 0.7053 & 23.96 & 0.1837  \\ 
0.05 & 0.01 & 0.7645 & 39.42 & 0.1691    \\
 & 0.02 & 0.9345 & 100.189 & 0.1448    \\     
\hline 
 & 0.0 & 0.6935 & 23.30 & 0.1733  \\  
0.10 & 0.01& 0.7505 & 37.88 & 0.1602 \\  
 & 0.02 & 0.9200 & 91.67 & 0.1382  \\ 
\hline 
 & 0.0& 0.6705 & 22.24 & 0.1566  \\   
0.15 & 0.01& 0.7390 & 35.50 &  0.1463  \\   
 & 0.02& 0.9010 & 79.92 & 0.1278  \\
\hline                 
\end{tabular}\label{E5}  
\end{table}

\begin{figure}[t!]
%\begin{subfigure}
\includegraphics[scale=0.32]{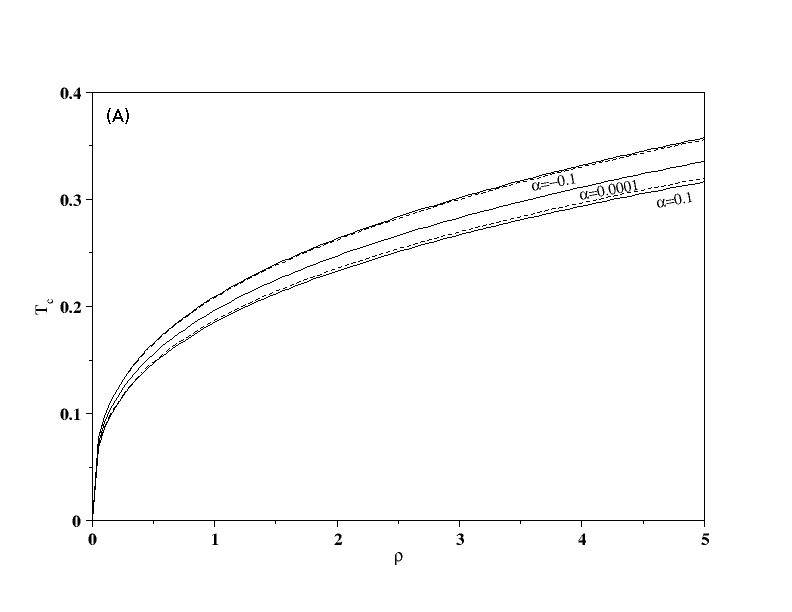} 
%\end{subfigure}
%\begin{subfigure}
\includegraphics[scale=0.4]{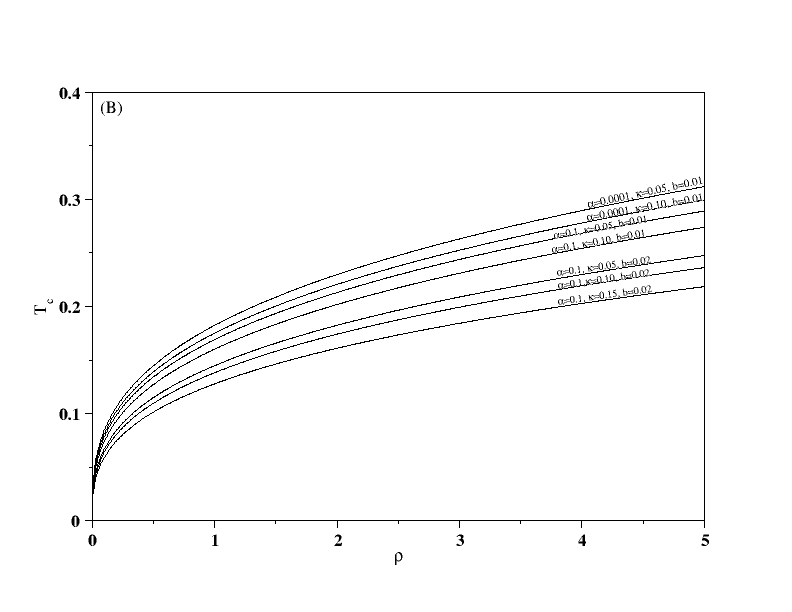}
%\end{subfigure}
%\includegraphics[scale=.4]{pic/debu2}
\caption{(A) $T_{c}$ vs. $\rho$ plot for Gauss-Bonnet holographic superconductors for different choice of parameters $\alpha$ (three sets) with the same value of $\kappa=0.0,~b=0$. The continuous curves correspond to numerical values whereas the dotted curves correspond to analytic values for $\alpha=-0.1,~\alpha=0.0001,~\alpha=0.1 $. Note that for $\alpha=0.0001$, the numerical and analytic curves are on top of each other.
(B) All curves correspond to analytic values for different choice of parameters $(\alpha,~\kappa,~b)$, namely, $(0.0001,~0.05,~0.01),~(0.0001,~0.10,~0.01),~(0.1,~0.05,~0.01),$
$~(0.1,~0.10,~0.01),~(0.1,~0.05,~0.02),~(0.1,~0.10,~0.02),$
$~(0.1,~0.15,~0.02)$(upper to lower).}
\end{figure}

%%%%%%%%%%%%%%%%%%%%%%%%%%%%%%%%%%%%%%%%%%%%%%%%%%%%%

\section{Condensation values and critical exponent}
In this section, we shall investigate the effect of BI coupling parameter and backreaction on the condensation operator near the critical temperature for Einstein and GB gravity. To proceed, we write down the field equation for $\phi (z)$ (\ref{e1aa}) near the critical temperature $T_{c}$ (using eq.(\ref{sol1}))  
\begin{eqnarray}
\phi^{\prime \prime}(z) - \frac{d-4}{z} \phi^{\prime}(z) + \frac{d-2}{r^2_{+}} b \phi^{\prime}(z)^{3} z^3 = \frac{\langle J\rangle ^{2}}{r^{2}_{+}} \mathcal{B}(z)\phi (z)
\label{cx1} 
\end{eqnarray}
where $\mathcal{B}(z)= \frac{2 z^{2\Delta_{+} -4}}{r_{+}^{2\Delta_{+} -4}}\frac{F^{2}(z)}{f(z)}\left( 1- \frac{b z^4}{r^2_{+}}\phi^{\prime}(z)^2 \right)^{\frac{3}{2}}.$ 
Note that we have kept the general form for the black hole spacetime ($f(z)$) which would be later set as the Einstein or the GB metric. We may now expand $\phi(z)$ in the small parameter $\frac{\langle J \rangle ^2}{r^{2}_{+}}$ as 
\begin{eqnarray}
\frac{\phi(z)}{r_{+}} = \lambda \left\{ (1-z^{d-3}) - \frac{b(\lambda^2|_{b=0}) (d-3)^3}{2(3d-7)} (1-z^{3d-7})\right\} + \frac{\langle J\rangle ^2}{r^{2}_{+}} \zeta (z)
\label{cx2} 
\end{eqnarray}
with $\zeta (1)= 0 =\zeta^{\prime}(1).$\\
Substituting eq.(\ref{cx2}) in eq.(\ref{cx1})and comparing the coefficient of $\frac{\langle J\rangle ^2}{r^{2}_{+}}$ on both sides of this equation (keeping terms upto $\mathcal{O}(b)$), we get the equation for the correction $\zeta(z)$ near the critical temperature
\begin{eqnarray}
\zeta^{\prime \prime}(z) -\left\{ \frac{d-4}{z} + 3b(\lambda^2|_{b=0}) (d-2)(d-3)^2 z^{2d-5} \right\} \zeta^{\prime}(z) = \lambda \frac{2 z^{2\Delta_{+} -4}}{r_{+}^{2\Delta_{+} -4}}\frac{F^{2}(z)}{f(z)}\mathcal{A}_{1} (z)
\label{cx3}
\end{eqnarray}
where $\mathcal{A}_{1} (z) = 1-z^{d-3} -\frac{3b(\lambda^2|_{b=0}) (d-3)^2}{2}\left\{(1-z^{d-3})z^{2d-4} +\frac{d-3}{3(3d-7)}(1-z^{3d-7}) \right\} $. \\
To solve this equation, we multiply it by $z^{-(d-4)} e^{\frac{3(d-2)(d-3)^2 b(\lambda^2|_{b=0})}{2d-4} z^{2d-4}} $ to get
\begin{eqnarray}
\frac{d}{dz}\left( z^{-(d-4)} e^{\frac{3(d-2)(d-3)^2 b(\lambda^2|_{b=0})}{2d-4} z^{2d-4}} \zeta^{\prime}(z) \right) = \lambda \frac{2 z^{2\Delta_{+} -2}}{r_{+}^{2\Delta_{+} -2}}\frac{z^{-(d-4)} F^{2}(z)}{g_{0}(z)+ 2\alpha g_{1}(z)}  e^{\frac{3(d-2)(d-3)^2 b(\lambda^2|_{b=0}) z^{2d-4}}{2d-4} } \mathcal{A}_{1} (z). 
\label{cx4}
\end{eqnarray}
Using the boundary conditions of $\zeta(z)$, we integrate the above equation between the limits $z=0$ and $z=1$. This leads to
\begin{eqnarray}
\frac{\zeta^{\prime}(z)}{z^{d-4}}\mid_{z\rightarrow 0} = -\frac{\lambda}{r^{2\Delta_{+} -2}_{+}} \mathcal{A}_{2}
\label{cx5}
\end{eqnarray}
where $\mathcal{A}_{2} = \int^{1}_{0} dz \frac{2 z^{2\Delta_{+}-2}z^{-(d-4)} F^{2}(z)}{(g_{0}(z)+ 2\alpha g_{1}(z))}  e^{\frac{3(d-2)(d-3)^2 b(\lambda^2|_{b=0})}{2d-4} z^{2d-4}} \mathcal{A}_{1} (z) $.\\
We now write down an interesting relation between $\zeta^{\prime}(z)$ and the $(d-3)$-th derivative of $\zeta(z)$ which we shall require in what follows
\begin{eqnarray}
\frac{\zeta^{(d-3)}(z=0)}{(d-4)!} = \frac{\zeta^{\prime}(z)}{z^{d-4}}|_{z\rightarrow 0}.
\label{cx6}
\end{eqnarray}
The asymptotic behaviour of $\phi(z)$ is given by eq.(\ref{bound1}). 
Eq.(\ref{cx2}) also gives the asymptotic behaviour of $\phi(z)$. Hence comparing these equations, we obtain 
\begin{eqnarray}
\mu -\frac{\rho}{r^{d-3}_{+}}z^{d-3} &=& \lambda r_{+} \left\{ (1-z^{d-3}) - \frac{b(\lambda^2|_{b=0}) (d-3)^3}{2(3d-7)} (1-z^{3d-7})\right\} \nonumber \\
&+& \frac{\langle J\rangle ^2}{r_{+}} \left\{\zeta(0)+z\zeta^{\prime}(0)+......+\frac{\zeta^{d-3}(0)}{(d-3)!} z^{d-3}+....\right\}.
\label{cx7}
\end{eqnarray}
Comparing the coefficient of $z^{d-3}$ on both sides of this equation, we get
\begin{eqnarray}
-\frac{\rho}{r^{d-3}_{+}} = -\lambda r_{+} + \frac{\langle J\rangle ^2}{r_{+}}\frac{\zeta^{d-3}(0)}{(d-3)!}~.
\label{cx8}
\end{eqnarray}
From eq.(s)(\ref{cx5}, \ref{cx6}) and eq.(\ref{cx8}), we obtain the relation between the charge density $\rho$ and the condensation operator $\langle J\rangle$ 
\begin{eqnarray}
\frac{\rho}{r^{d-2}_{+}} = \lambda\left[1 + \frac{\langle J\rangle ^2}{r^{2\Delta_{+}}_{+}}\frac{\mathcal{A}_{2}}{(d-3)} \right].
\label{cx9}
\end{eqnarray}  
Using eq.(\ref{de9}) and the definition of $\lambda$, we simplify
eq.(\ref{cx9}) to get
\begin{eqnarray}
\langle J\rangle ^2 = \frac{(d-3)(4\pi T_{c})^{2\Delta_{+}}}{\mathcal{A}_{2}[(d-1)- \frac{(d-3)^2}{(d-2)}\kappa^2_{i} (\lambda^2|_{\kappa_{i-1}})]^{2\Delta_{+}}}\left(\frac{T_{c}}{T}\right)^{d-2} \left[1- \left(\frac{T}{T_{c}}\right)^{d-2} \right].
\label{cx10}
\end{eqnarray}
%The temperature is away from (but close to) the critical temperature (i.e. $T \approx T_{c}$) and therefore we have
%\begin{eqnarray}
%\left(\frac{T_{c}}{T}\right)^{d-2} \left[1- \left(\frac{T}{T_{c}}\right)^{d-2} \right] &=& \left(\frac{T_{c}}{T}\right)^{d-2}\left[1- \left(\frac{T}{T_{c}}\right)\right]\left[1+\frac{T}{T_{c}} + \left(\frac{T_{c}}{T}\right)^2 +.....+\left(\frac{T_{c}}{T}\right)^{d-3} \right] \nonumber \\
%&=& (d-2)\left[1- \left(\frac{T}{T_{c}}\right)\right]
%\label{cx11}
%\end{eqnarray} 
From this we finally obtain the relation between the condensation operator and the critical temperature in $d$-dimension for $T\rightarrow T_c$
\begin{eqnarray}
\langle J\rangle = \beta T^{\Delta_{+}}_{c} \sqrt{1-\frac{T}{T_{c}}}
\label{cx12}
\end{eqnarray}
where $\beta = \sqrt{\frac{(d-3)(d-2)}{\mathcal{A}_{2}}} \left[\frac{4\pi}{(d-1)- \frac{(d-3)^2}{(d-2)}\kappa^2_{i} (\lambda^2|_{\kappa_{i-1}})}\right]^{\Delta_{+}}.$\\

\noindent Once again we find that the critical exponent is $1/2$
which agrees with the universal mean field value.  
We shall now set $d=5$ and $m^2=-3$ for the rest of our analysis. The choice for $m^2$ yields $\Delta_{+}=3$. Eq.(\ref{cx12}) now simplifies to 
\begin{eqnarray}
\langle J\rangle = \beta T^{3}_{c} \sqrt{1-\frac{T}{T_{c}}}~.
\label{cx13}
\end{eqnarray}
The expressions for $\mathcal{A}_{1}(z)$, $\mathcal{A}_{2}$ and $\beta$ simplify to 
\begin{eqnarray}
\mathcal{A}_{1} (z) &=& 1-z^2 -\frac{b(\lambda^2|_{b=0})}{2}(1-z^2)\left[12 z^6 + \frac{1-z^8}{1-z^2}\right] \nonumber \\ 
&=& (1-z^2)\left[1-\frac{b(\lambda^2|_{b=0})}{2} (1+z^2 +z^4 +13z^6) \right] \nonumber \\
\mathcal{A}_{2} &=& \int^{1}_{0} dz \frac{2 z^{3} F^{2}(z)}{[g_{0}(z)+ 2\alpha g_{1}(z)]}  e^{6 b(\lambda^2|_{b=0}) z^{6}} \mathcal{A}_{1}(z) \nonumber\\
\beta &=& \sqrt{\frac{6}{\mathcal{A}_{2}}} \left[\frac{\pi}{1- \frac{1}{3}\kappa^2_{i} (\lambda^2|_{\kappa_{i-1}} )} \right]^{3}~.
\label{cx14}
\end{eqnarray}
Simplifying $\mathcal{A}_{2} $ upto $\mathcal{O}(b)$, we obtain  
\begin{eqnarray}
\mathcal{A}_{2} &=& \int^{1}_{0} dz \frac{2 z^{3} F^{2}(z) (1-z^2)}{[g_{0}(z)+ 2\alpha g_{1}(z)]} \left\{1- \frac{b(\lambda^2|_{b=0})}{2} (1+z^2 +z^4 +z^6) \right\}
\end{eqnarray}
In Einstein gravity, the metric term should be $g_{0}(z)$ (because $\alpha = 0$). Using eq.(s)(\ref{metr33}, \ref{eq50}) and computing $\mathcal{A}_{2}$ with $\tilde{\alpha}= 0.7218$ for $\kappa=0, ~b=0$, we obtain $\beta= 238.908$ which is in very good agreement with the exact result $\beta= 238.958$ \cite{hs24}.\\
Now we shall proceed to include the effect of the BI parameter $(b\neq 0)$ and backreaction $(\kappa \neq 0)$ in our analysis. For $\kappa =0, ~b=0.01$, computing $\mathcal{A}_{2}$ with $\tilde{\alpha}= 0.7540$, we get $\beta= 270.834$ which agrees wonderfully with the exact result $\beta= 271.612$ \cite{hs24}. For $\kappa =0.05, ~b=0$, computing $\mathcal{A}_{2}$ with $\tilde{\alpha}= 0.7195$, we get $\beta= 248.959$. In Table \ref{cxEt1}, we present the analytic results for Einstein gravity.

\begin{table}[h!]
\caption{The analytical results for the condensation operator and the critical temperature with backreaction and Born-Infeld parameter in Einstein gravity ($\alpha=0$)}   
\centering                          
\begin{tabular}{c c c c c c c}            
\hline                       
$\kappa$& $b$ & $\tilde{\alpha} $ & $\lambda^{2}_{SL}$  & $\mathcal{A}_{2}$ & $\beta |_{SL} =\frac{\langle J\rangle}{T^3_{c} \sqrt{1-T/T_{c}}}$ \\
\hline
0.0 & 0.0 & 0.7218 & 18.23 & 0.101062 & 238.908  \\
0.0 & 0.01 & 0.7540 & 25.91 & 0.07864 & 270.834 \\
\hline
0.05 & 0.0 & 0.7195 & 18.11 & 0.10202 & 248.959\\ 
0.05 & 0.01 & 0.7525 & 25.68 & 0.07938 & 287.815 \\
\hline 
0.10 & 0.0 & 0.7122 & 17.75 & 0.10508 & 282.416 \\
0.10 & 0.01 & 0.7455 & 25.02 & 0.08203 & 346.824 \\ 
\hline              
\end{tabular}
\label{cxEt1}  
\end{table}

\noindent In Gauss-Bonnet gravity, we use eq.(s)(\ref{metr20z}, \ref{metr41}) for the form of the metric. We set the GB parameter $\alpha= 0.1$ and $\kappa=0$. Computing $\mathcal{A}_{2}$ with $\tilde{\alpha}= 0.7080$ for $b=0$, we obtain $\beta= 244.112$ which agrees very well with the exact result $\beta= 243.897$ \cite{hs24}. For $b=0.01, ~\tilde{\alpha}= 0.7665$, we obtain $\beta= 294.147$ which is once again in good agreement with exact $\beta= 290.107$ \cite{hs24}. In Table \ref{cxEt2}, we present the analytic results for the condensation operator for GB gravity. \\

\noindent In figure 3, the plot of $\frac{\langle J\rangle}{T^{3}_{c}}$ vs. $\frac{T}{T_{c}}$ is shown for Einstein gravity and GB gravity for different choices of $\kappa, ~b$.    

\begin{table}[t]
\caption{The analytical results for the condensation operator and the critical temperature with backreaction and Born-Infeld parameter in GB gravity ($\alpha=0.1$)}   
\centering                          
\begin{tabular}{c c c c c c c}            
\hline                       
$\kappa$& $b$ & $\tilde{\alpha} $ & $\lambda^{2}_{SL}$  & $\mathcal{A}_{2}$ & $\beta |_{SL} =\frac{<J>}{T^3_{c} \sqrt{1-T/T_{c}}}$ \\
\hline
0.0 & 0.0 & 0.7080 & 24.18 & 0.096798 & 244.112 \\
0.0 & 0.01 & 0.7665 & 39.96 & 0.066669 & 294.147 \\
\hline
0.05 & 0.0 & 0.7053 & 23.96 & 0.098019 & 257.863\\ 
0.05 & 0.01 & 0.7645 & 39.42 & 0.067623 & 323.298 \\
\hline 
0.10 & 0.0 & 0.6935 & 23.30 & 0.10255 & 304.436 \\
0.10 & 0.01 & 0.7505 & 37.88 & 0.071655 & 432.956 \\ 
\hline              
\end{tabular}
\label{cxEt2}  
\end{table}

\begin{figure}[t!]
%\begin{subfigure}
\includegraphics[scale=0.30]{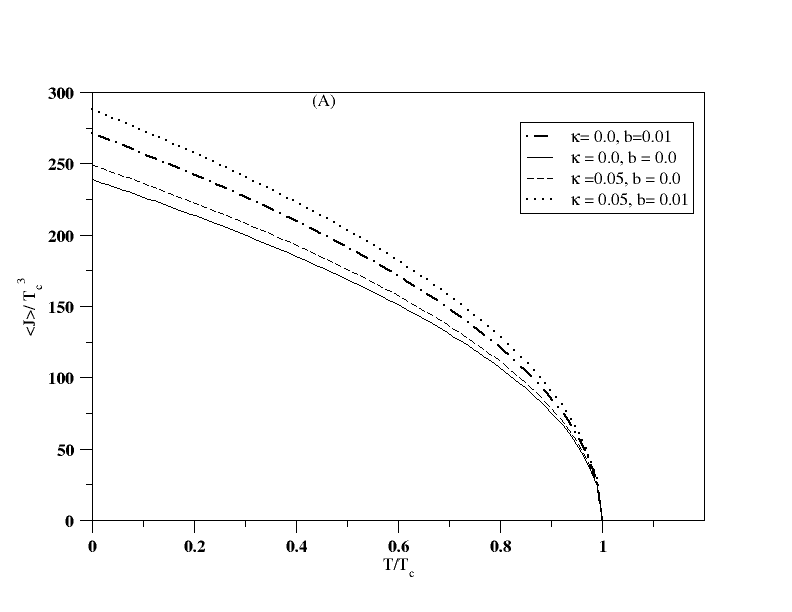} 
%\end{subfigure}
%\begin{subfigure}
\includegraphics[scale=0.30]{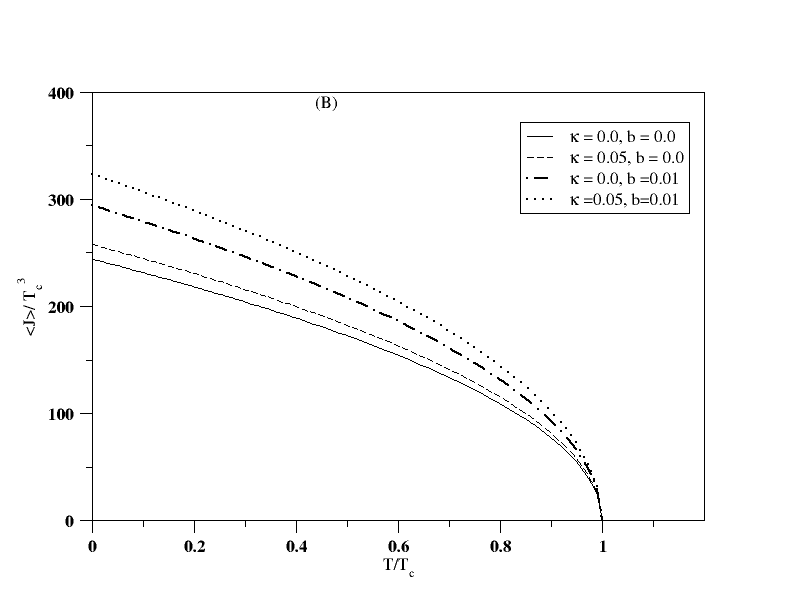}
%\end{subfigure}
%\includegraphics[scale=.4]{pic/debu2}
\caption{(A) $\langle J\rangle/T^{3}_{c}$ vs. $T/T_{c}$ plot for Einstein holographic superconductors for different choices of parameters $\kappa,~b$.
(B)  $\langle J\rangle/T^{3}_{c}$ vs. $T/T_{c}$ plot for Gauss-Bonnet holographic superconductors with GB parameter $\alpha=0.1$ for different choices of parameters $\kappa,~b$.}
\end{figure}

%%%%%%%%%%%%%%%%%%%%%%%%%%%%%%%%%%%%%%%%%%%%%%%%%%%%%%%%%%%%%%%%%%%%%%%%%%%%%%%%%%%%%%%%%%%%%%%%%%%%%%
%%%%%%%%%%%%%%%%%%%%%%%%%%%%%%%%%%%%%%%%%%%%%%%%%%%%%%%%%%%%%%%%%%%%%%%%%%%%%%%%%%%%%%%%%%%%%%%%%%%%%%

\section{Conclusions}
In this paper, we have analytically calculated the relation between the critical temperature and the charge density of higher dimensional holographic superconductors in the framework of Born-Infeld electrodynamics taking into account the effect of backreaction of the matter fields on the spacetime metric. In particular the relation between the critical temperature and the charge density holds for a $d$-dimensional holographic superconductor and is one of the main result in this paper. We observe that the condensation gets hard to form in the presence of the Born-Infeld parameter. It is also noted that the condensate gets harder to form in Gauss-Bonnet gravity than Einstein gravity in $4+1$-dimensions. The inclusion of the effect of backreaction of the matter fields on the spacetime metric makes the condensate even harder to form. We find that our results are in very good agreement with the existing numerical results \cite{liu, yao}. We also derive an expression for the condensation operator in $d$-dimensions and then analyse the effects of the Born-Infeld and Gauss-Bonnet parameters in the presence of back reactions in $d=5$ dimensions. Our results agree wonderfully with the available numerical results in the literature.
The mean field value of $1/2$ for the critical exponent is obtained in our analysis.

We would now like to mention the importance of our results obtained analytically. 
%The consistency between the analytical and numerical results indicates that the Sturm-Liouville method is a powerful analytic approach to investigate holographic superconductors even in the presence of backreaction.
It is evident that the Sturm-Liouville eigenvalue method is a powerful analytic approach to investigate holographic superconductors taking into account the effect of various parameters, namely, the Born-Infeld parameter and the Gauss-Bonnet coupling parameter. One of the great advantages of this approach is that it is also found to be applicable away from the probe limit. This can be inferred by comparing the analytical results with the numerical results.  
It should also be appreciated that the analytical method is always more reliable than the numerical approach since the reliability of the numerical results decreases
when the temperature $T$ approaches to zero \cite{hs3, siop}.
We further point out that our analytical results obtained by the Sturm-Liouville eigenvalue method also agree with the results obtained from an alternative analytic technique known as the matching method \cite{hs8}, \cite{yao}, \cite{sgan}. Our general result presented in $d$-dimensions can also be applied for values of $d\geq 4$ for Einstein gravity and $d>5$ for Gauss-Bonnet gravity.
Work in the future direction is in progress where we would like to analyze the same set up immersed in an external magnetic field \cite{johnson}.

\section*{Acknowledgments} DG would like to thank S.N.~Bose National Centre for Basic Sciences, Kolkata, India for providing facilities and DST-INSPIRE for financial support. 
S.G. acknowledges the support by DST SERB under Start Up Research Grant (Young Scientist), File No.YSS/2014/000180.

%%%%%%%%%%%%%%%%%%%%%%%%%%%%%%%%%%%%%%%%%%%%%%%%%%


\begin{thebibliography}{99}
\baselineskip=0.6 cm
\bibitem{bcs} J. Bardeen, L. N. Cooper, J. R. Schrieffer, ``Theory of Superconductivity", Phys. Rev. 108, 1175 (1957). 
%\bibitem{parks}R.D. Parks, {\it{Superconductivity}}, Marcel Dekker Inc., New York U.S.A. (1969).
\bibitem{adscft1} J. M. Maldacena, ``The Large N Limit of Superconformal Field Theories and Supergravity", Adv. Theor. Math. Phys. 2, 231 (1998).
\bibitem{adscft2} E. Witten, ``Anti De Sitter Space And Holography", Adv. Theor. Math. Phys. 2, 253 (1998).
\bibitem{adscft3} S.S. Gubser, I.R. Klebanov, A.M. Polyakov, ``Gauge Theory Correlators from Non-Critical String Theory", Phys. Lett. B 428, 105 (1998).
\bibitem{adscft4} O. Aharony, S.S. Gubser, J.M. Maldacena, H. Ooguri, Y. Oz, ``Large N Field Theories, String Theory and Gravity", Phys. Rept. 323, 183 (2000).
\bibitem{hs1} S.S. Gubser, ``Phase transitions near black hole horizons", Class. Quant. Grav.  22, 5121 (2005).
\bibitem{hs2} S.S. Gubser, ``Breaking an Abelian gauge symmetry near a black hole horizon", Phys. Rev. D 78, 065034 (2008).
\bibitem{hs3} S.A. Hartnoll, ``Lectures on holographic methods for condensed matter physics", Class. Quantum Grav. 26, 224002 (2009).
\bibitem{nw1} S.-S. Lee, ``A Non-Fermi Liquid from a Chared Black Hole : A Critical Fermi Ball", Phys. Rev. D 79, 086006 (2009).
\bibitem{nw2} H. Liu, J. McGreevy, D. Vegh, ``Non-Fermi liquids from holography", Phys. Rev. D 83, 065029 (2011).
\bibitem{nw3} T. Nishioka, S. Ryu, T. Takayanagi, ``Holographic Superconductor/Insulator Transition at Zero Temperature", JHEP 1003, 131 (2010). 
\bibitem{hs4} C.P. Herzog, ``Lectures on holographic superfluidity and superconductivity", J. Phys. A  42, 343001 (2009).
\bibitem{hs5} G.T. Horowitz, ``Introduction to Holographic Superconductors", arXiv:1002.1722 [hep-th].
\bibitem{hs6} S.A. Hartnoll, C.P. Herzog, G.T. Horowitz, ``Building a Holographic Superconductor", Phys. Rev. Lett. 101, 031601 (2008).
\bibitem{siop} G. Siopsis, J. Therrien, ``Analytic calculation of properties of holographic superconductors", JHEP 05 (2010) 013.
\bibitem{hs8} R. Gregory, S. Kanno, J. Soda, ``Holographic Superconductors with Higher Curvature Corrections", JHEP 0910 (2009) 010.
\bibitem{hs9} S. A. Hartnoll, C. P. Herzog, G. T. Horowitz, ``Holographic superconductors", JHEP 12, 015 (2008).
\bibitem{hs9a} H.B. Zeng, X. Gao, Y. Jiang, H.-S. Zong, ``Analytical Computation of Critical Exponents in Several Holographic Superconductors", JHEP 05 (2011) 002.
\bibitem{hs9b} H.F. Li, R.-G. Cai, H.-Q. Zhang, ``Analytical Studies on Holographic Superconductors in Gauss-Bonnet Gravity", JHEP 04 (2011) 028.
\bibitem{hs9c} R.G. Cai, H.-F. Li, H.-Q. Zhang, ``Analytical studies on holographic insulator/superconductor phase transitions", Phys. Rev. D 83 (2011) 126007.
\bibitem{hs9d} S.S. Gubser, S.S. Pufu, ``The gravity dual of a p-wave superconductor", JHEP 11 (2008) 033.
\bibitem{hs14}  Q. Y. Pan, B. Wang, E. Papantonopoulos, J. Oliveira, A. Pavan, ``Holographic superconductors with various condensates in Einstein-Gauss-Bonnet gravity", Phys. Rev. D 81,106007 (2010).
\bibitem{hs15}  R.G. Cai, H. Zhang, ``Holographic superconductors with Horava-Lifshitz black holes", Phys. Rev. D 81, 066003 (2010).
\bibitem{hs16}G. T. Horowitz, M. M. Roberts, ``Holographic superconductors with various condensates", Phys. Rev. D 78, 126008 (2008).
\bibitem{hs17} G. T. Horowitz, M. M. Roberts, ``Zero Temperature Limit of Holographic Superconductors", JHEP  0911 (2009) 015.
\bibitem{hs18} Q Pan, J. Jing, B. Wang, ``Analytical investigation of the phase transition between holographic insulator and superconductor in Gauss-Bonnet gravity", JHEP 1111 (2011) 088.
\bibitem{hs19} J. Jing, S. Chen, ``Holographic superconductors in the Born–Infeld electrodynamics", Phys. Lett. B 686 (2010) 68.
\bibitem{hs20} J. Jing, Q Pan, S. Chen, ``Holographic superconductors with Power-Maxwell field", JHEP 1111 (2011) 045. 
\bibitem{hs21} J. Jing, Q Pan,  B. Wang, ``Holographic superconductor models with the Maxwell field strength corrections", Phys. Rev. D 84,126020, (2011).
\bibitem{hs22} J. Jing,  L. Wang, Q Pan, S. Chen, ``Holographic superconductors in Gauss-Bonnet gravity with Born-Infeld electrodynamics", Phys. Rev. D 83,066010, (2011).
\bibitem{sgdc1} S.~Gangopadhyay, D.~Roychowdhury, ``Analytic study of properties of holographic superconductors in Born-Infeld electrodynamics", JHEP 05 (2012) 002.
\bibitem{hs24}S.~Gangopadhyay, D.~Roychowdhury, ``Analytic study of Gauss-Bonnet holographic superconductors in Born-Infeld electrodynamics", JHEP 05 (2012) 156.
\bibitem{rb} R. Banerjee, S.~Gangopadhyay, D.~Roychowdhury, A. Lala, ``Holographic s-wave condensate with nonlinear electrodynamics: A nontrivial boundary value problem", Phys. Rev. D 87 (2013) 104001.
\bibitem{sgm} S. Gangopadhyay, ``Holographic superconductors in Born–Infeld electrodynamics and external magnetic field", Mod. Phys. Lett. A 29 (2014) 1450088.
%\bibitem{sgdc1} S.~Gangopadhyay, D.~Roychowdhury, ``Analytic study of properties of holographic superconductors in Born-Infeld electrodynamics", JHEP 05 (2012) 002.
\bibitem{deser} D.G. Boulware, S. Deser, ``String-generated gravity models", Phys. Rev. Lett. 55 (1985) 2656. 
\bibitem{wheeler} J.T. Wheeler, ``Symmetric solutions to the Gauss-Bonnet extended Einstein equations", Nucl. Phys. B 268 (1986) 737.
\bibitem{cai} R.G.~Cai, ``Gauss-Bonnet black holes in AdS spaces", Phys. Rev. D 65 (2002) 084014. 
%\bibitem{hs24}S.~Gangopadhyay, D.~Roychowdhury, ``Analytic study of Gauss-Bonnet holographic superconductors in Born-Infeld electrodynamics", JHEP 05 (2012) 156.
\bibitem{nell}S.S. Gubser, A. Nellore, ``Low-temperature behavior of the Abelian Higgs model in anti-de Sitter space", JHEP 0904 (2009) 008.
\bibitem{russo}F.~Aprile, J.G. Russo, ``Models of Holographic superconductivity", Phys. Rev. D 81 (2010) 026009.
\bibitem{siani}M. Siani, ``Holographic superconductors and higher curvature corrections", JHEP 1012 (2010) 035.
\bibitem{greg} R. Gregory, ``Holographic superconductivity with Gauss-Bonnet gravity", J. Phys. Conf. Ser. 283 (2011) 012016.
\bibitem{sgplb}S.~Gangopadhyay, ``Analytic study of properties of holographic superconductors away from the probe limit", Phys. Lett. B 724 (2013) 176.
\bibitem{yao} W. Yao, J. Jing, ``Analytical study on holographic superconductors for Born-Infeld electrodynamics in Gauss-Bonnet gravity with backreactions", JHEP 05 (2013) 101. 
\bibitem{chen} Q. Pan, J. Jing, B. Wang, S. Chen, ``Analytical study on holographic superconductors with backreactions", JHEP 06 (2012) 087.
\bibitem{betti} Y. Brihaye, B. Hartmann, ``Holographic superconductors in $3+1$ dimensions away from the probe limit", Phys. Rev. D 81, 126008 (2010).
\bibitem{liu} Y. Liu, Y. Peng, B. Wang, ``Gauss-Bonnet holographic superconductors in Born-Infeld electrodynamics with backreactions", arXiv:1202.3586.
\bibitem{sgan}S.~Gangopadhyay, D.~Roychowdhury, ``Analytic study of properties of holographic $p$-wave superconductors", JHEP 08 (2012) 104.


%\bibitem{hs16}G. T. Horowitz, M. M. Roberts, Phys. Rev. D 78, 126008 (2008).

%\bibitem{gub} S. S. Gubser, A. Nellore, JHEP 04, 008 (2009).
%\bibitem{aprile} F. Aprile, J.G. Russo, Phys. Rev. D 81 (2010) 026009.
%\bibitem{liu} Y. Liu, Y.W. Sun, JHEP 07, 099 (2010).
%\bibitem{barc} L. Barclay, R. Gregory, S. Kanno, P. Sutcliffe, JHEP 12, 029 (2010).
%\bibitem{siani} M. Siani, JHEP 12, 035 (2010).
%\bibitem{gregory} R. Gregory, J. Phys. Conf. Ser. 283 (2011) 012016.
%\bibitem{pan} Q. Pan, J. Jing, B. Wang, S. Chen, JHEP 06, 087 (2012).

%\bibitem{hs13} S. S. Gubser, S. S. Pufu, JHEP  11, 033 (2008).
%\bibitem{hs13b} S.S. Gubser, Phys. Rev. Lett. 101 (2008) 191601.
%\bibitem{hs13a}M.M. Roberts, S.A. Hartnoll, JHEP 08 (2008) 035.
%\bibitem{hs7} G. Siopsis, J. Therrien, JHEP 05, 013 (2010).


%\bibitem{hs10} H. B. Zeng, X. Gao, Y. Jiang, H. S. Zong, JHEP 1105 (2011) 002.
%\bibitem{hs11} H. F. Li, R. G. Cai, H. Q. Zhang, JHEP 1104 (2011) 028.
%\bibitem{hs12} R. G. Cai, H. F. Li, H. Q. Zhang, Phys. Rev. D 83, 126007 (2011).
%\bibitem{hs14}  Q. Y. Pan, B. Wang, E. Papantonopoulos, J. Oliveira, A. Pavan, Phys. Rev. D 81,106007 (2010).
%\bibitem{hs15}  R.G. Cai, H. Zhang, Phys. Rev. D 81, 066003 (2010).

%\bibitem{hs17} G. T. Horowitz, M. M. Roberts, JHEP  0911 (2009) 015.
%\bibitem{hs181}  Q. Pan, J. Jing, B. Wang, S. Chen ,JHEP 06 (2012) 087.
%\bibitem{hs19} J. Jing, S. Chen, Phys. Lett. B 686 (2010) 68.
%\bibitem{hs20} J. Jing, Q Pan, S. Chen, JHEP 1111 (2011) 045. 
%\bibitem{hs21} J. Jing, Q Pan,  B. Wang, Phys. Rev. D 84,126020, (2011).
%\bibitem{hs22} J. Jing,  L. Wang, Q Pan, S. Chen, Phys.Rev.D 83,066010, (2011).
%\bibitem{hs23}S.~Gangopadhyay, D.~Roychowdhury, JHEP 05 (2012) 002. 
%\bibitem{sgan}S.~Gangopadhyay, D.~Roychowdhury, JHEP 08 (2012) 104.
%\bibitem{sgplb}S.~Gangopadhyay, Phys. Lett. B 724 (2013) 176.
%\bibitem{jhep05-2013}

\bibitem{johnson} T. Albash, C.V. Johnson, ``A holographic superconductor in an external magnetic field", JHEP 09 (2008) 121.
%\bibitem{BI} M. Born, L. Infeld, Proc. Roy. Soc. A 144, 425 (1934).
%\bibitem{plb2} E. S. Fradkin, A. A. Tseytlin, Phys. Lett. B 163, 123 (1985).
%\bibitem{nplb1} A. A. Tseytlin, Nucl. Phys. B 276, 391 (1986).
%\bibitem{tamaki} T. Tamaki, T. Torii, Phys. Rev. D 62, 061501 (2000).
%\bibitem{plb1} Mauricio Cataldo, Alberto Garcia, Phys. Lett. B 456, 28–33 (1999).
%\bibitem{cqg2} G. W. Gibbons, C. A. R. Herdeiro, Class. Quantum Grav. 18, 1677 (2001).
%\bibitem{plb3} A. Sheykhi, Phys. Lett. B 662, 7 (2008).
%\bibitem{jhep} R. G. Cai, Y. W. Sun, JHEP 09, 115 (2008).
%\bibitem{fernandogrg} S. Fernando, D. Krug, Gen. Relativ. Gravit. 35, 129 (2003).
%\bibitem{fernando} S. Fernando, Phys. Rev. D 74 , 104032 (2006).
%\bibitem{cai} R. G. Cai, Da-Wei Pang, A. Wang, Phys. Rev. D 70, 124034 (2004).
%\bibitem{myung} Y. S. Myung, Y. W. Kim, Y. J. Park, Phys. Rev. D 78 , 084002 (2008).
%\bibitem{dey} T. K. Dey, Phys. Lett. B 595, 484-490 (2004).
%\bibitem{olivera} O. Miscovic, R. Olea, Phys. Rev. D 77, 124048 (2008).
%\bibitem{bf1}P. Breitenlohner, D. Z. Freedman, Phys. Lett. 115B, (1982) 197.
%\bibitem{bf2}P. Breitenlohner, D. Z. Freedman, Ann. Phys. 144 (1982) 197.

\end{thebibliography}
\end{document}